\documentclass[onecolumn]{article}
\usepackage{graphicx}
\usepackage{here,amsmath,cases,amssymb,bm,subfigure,supertabular}

 \evensidemargin \oddsidemargin

\title{
\LARGE \bf
Noise-Induced Spatial Pattern Formation in Stochastic Reaction-Diffusion
Systems
\footnote{
\copyright 2012 IEEE. Personal use of this material is
permitted. Permission from IEEE must be obtained for all other uses, in
any current or future media, including reprinting/republishing this
material for advertising or promotional purposes, creating new
collective works, for resale or redistribution to servers or lists, or
reuse of any copyrighted component of this work in other works. 
The citation of this work should be as follows: Y. Hori and S. Hara,
``Noise-Induced Spatial Pattern Formation in Stochastic
Reaction-Diffusion Systems,''  {\it Proceedings of IEEE Conference on 
Decision and Control}, pp. 1053--1058, 2012.
}
}

\author{Yutaka Hori \and Shinji Hara \footnote{Y. Hori and S. Hara are
with Department of Information Physics and Computing, The
 University of Tokyo, 7-3-1 Hongo, Bunkyo-ku, Tokyo 113-8656,
 Japan. \{Yutaka\_Hori, Shinji\_Hara\}@ipc.i.u-tokyo.ac.jp}  % Please supply
}
\date{}

\begin{document}
\maketitle

%%%%%%%%%%%%%%%%%%%%%%%%%%%%%%%%%%%%%%%%%%%%%%%%%%%%%%%%%%%%%%%%%%%%%%%%%%%%%%%%

\begin{abstract}
This paper is concerned with stochastic reaction-diffusion
kinetics governed by the reaction-diffusion master
equation. Specifically, the primary goal of this paper is to
provide a mechanistic basis of Turing pattern formation that
is induced by intrinsic noise. To this end, we first derive an
approximate reaction-diffusion system by using linear noise
approximation. We show that the approximated system has a
certain structure that is associated with a coupled dynamic
multi-agent system. 
This observation then helps us derive
an efficient computation tool to examine the spatial power
spectrum of the intrinsic noise. We numerically demonstrate
that the result is quite effective to analyze noise-induced Turing
pattern. Finally, we illustrate the theoretical mechanism behind the 
noise-induced pattern formation with a $\mathcal{H}_2$ norm 
interpretation of the multi-agent system.
\end{abstract}

%%%%%%%%%%%%%%%%%%%%%%%%%%%%%%%%%%%%%%%%%%%%%%%%%%%%%%%%%%%%%%%%%%%%%%%%%%%%%%%%
\section{Introduction}
The auto-regulation mechanism of the biological pattern formation 
is a long standing question in biology. 
It was Turing \cite{Turing1952} who first presented a mathematical model 
that possibly accounts for the pattern formation in embryonic development.
Specifically, %he mathematical model reaction-diffusion
it was shown that a pair of reaction-diffusion equations can
spontaneously exhibit spatial structure after a small perturbation to a spatially homogeneous
equilibrium. 
Nowadays, such patterns are called Turing pattern, and 
analysis ranges to a wide variety of biological applications (see \cite{Kondo2010} for example).

% model that accounts for 
% a simple reaction-diffusion model consisting of two molecular species 
% can account for the generation of biological patterns. 
% That is, small disturbance to the spatially homogeneous equilibrium 
% can lead to complex spatial structure. 
%-Turing pattern is well-known. has been analyzed so much.
%-- slightly perturb around homogeneous equilibrium

\par
\smallskip
Although many existing studies rely on deterministic
reaction-diffusion models, intrinsic noise
is often a unignorable factor leading to a drastic change of the dynamics
\cite{Elowitz2002}. 
% In particular, the intrinsic noise is due to the chemical reactions 
% under the small copy number of molecules, thus is more important 
%  for biochemical reactions.
It is known that the dynamics of stochastic chemical reactions in a cell
 follows the chemical master equation (CME) \cite{Gillespie1992}. 
A standing assumption of the CME is that reactants are well-mixed in a
 cell.
However, recent studies revealed that molecules can localize
inside a cell, and the localization plays an important role, for example, in 
cell division \cite{Rudner2010}. % and carbon fixation process \cite{*****}. 
Hence, it is important to analyze the stochastic dynamics of 
 biochemical reactions under the spatially inhomogeneous environment.

\par
\smallskip
The reaction-diffusion master equation (RDME) \cite{Gardiner1976} 
describes the dynamics of the molecular copy numbers 
under the stochastic reaction-diffusion process.
In this formulation, the spatial domain is partitioned into small voxels 
so that the molecules are well-mixed inside each voxel, but the copy numbers
of molecules are small enough to preserve the stochasticity. Thus, the
dynamics of reactions inside each voxel is governed by the chemical master
equation (CME) \cite{Gillespie1992}, while the diffusion is modeled as 
the exchange of the molecules between voxels.

%-stochastic is important
%--RDME mesoscopic
\par
\smallskip
In \cite{Biancalani2010}, the stochastic pattern formation was studied for 
Brusselator \cite{Glandsdorff1971} based on the RDME. 
Interestingly, it was observed that 
Brusselator can exhibit spatial patterns even when the deterministic
reaction-diffusion model converges to a homogeneous equilibrium. 
This implies that intrinsic fluctuation drives a particular spatial
mode, and results in inhomogeneous spatial structure. 
This research thereafter inspired the investigation of 
the noise-induced Turing patterns in Levin-Segel model \cite{Butler2011} 
and the noise-induced spatio-temporal oscillations in Brusselator \cite{Biancalani2011}.

\par
\smallskip
In these papers \cite{Biancalani2010, Butler2011, Biancalani2011}, an approximate model obtained by linear noise
approximation (LNA) \cite{vanKampen2007} was used to compute the spatial power
spectrum of the patterns.
% is useful to analyze the spatial power
%spectrum of the stochastic patterns. 
Later, the LNA for reaction-diffusion systems was 
 formulated in a more general form in \cite{Scott2011}, 
 and the analytic results obtained by the approximation agreed with the
 stochastic simulations.
Although these methods were shown to give a good approximation, 
the mechanistic basis of the stochastic pattern formation is not
 necessarily revealed. 
Hence, it is desirable that the system be analyzed from a system
 theoretic viewpoint to understand the mechanism behind the phenomenon.

%Although we observed that the effectiveness of the LNA, 
%intuitive understanding were obtained from 
%numerical simulations were presented to show the 

\par
\smallskip
The goal of this paper is to provide a mechanistic understanding 
of the stochastic reaction-diffusion system, and to reveal
the mechanism of the noise-induced Turing pattern from a control theoretic
viewpoint.
Specifically, it is shown that the computation of the covariance of intrinsic
noise can be viewed as the $\mathcal{H}_2$ norm computation of a coupled 
dynamic multi-agent system, where disturbance inputs are injected to 
each agent's states and sensed outputs.
Then, an efficient method to compute the spatial power spectrum
 is derived by using the characteristic structure of 
the reaction-diffusion system. 
Using these analytic tools, we analyze noise-induced Turing patterns 
observed in the celebrated Gray-Scott model \cite{Gray1984, Pearson1993}. 
It should be remarked that to the authors' knowledge, this paper gives 
the first presentation of the noise-induced Turing pattern in the
Gray-Scott model \cite{Gray1984, Pearson1993}. 
Finally, we reveal the mechanism of the noise-induced pattern
formation with the $\mathcal{H}_2$ norm interpretation of the multi-agent system.

\par
\smallskip
This paper is organized as follows.
In the next section, the RDME is briefly presented. 
Then, an approximated system of the RDME is derived by using
the LNA in Section III.
In Section IV, the characteristic structure of the approximated
system is revealed, and its control theoretic interpretation is
presented. Furthermore, an efficient method to compute spatial 
power spectrum is obtained.
In Section V, noise-induced Turing patters are demonstrated with 
the Gray-Scott model, and their mechanism is illustrated.
Finally, the paper is concluded in Section VI.

\section{Model description of stochastic reaction-diffusion systems}
\begin{figure}
\centering
\includegraphics[clip,width=12.3cm]{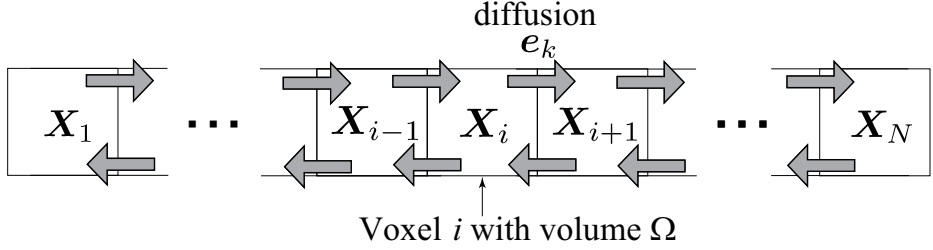}
\caption{The reaction-diffusion scheme considered in the RDME formulation.}
\label{scheme-fig}
\end{figure}

In this section, the dynamics of stochastic reaction-diffusion system, 
which we call reaction-diffusion master equation (RDME) \cite{Gardiner1976} is introduced. 

\par
\smallskip
Consider a set of chemical reactions that consists of 
$n$ molecular species, $\mathcal{M}_1, \mathcal{M}_2, \cdots, \mathcal{M}_n$, and $m_0$
reactions, $\mathcal{R}_{1}, \mathcal{R}_2, \cdots, \mathcal{R}_{m_0}$,  in a spatial domain.
Suppose the domain is partitioned into $N$ voxels,
$\mathcal{V}_1, \mathcal{V}_2, \cdots, \mathcal{V}_N$ with the
same volume $\Omega$. 
It is assumed that molecules are well-mixed within each voxel, 
and can react with those in the same voxel.
Figure \ref{scheme-fig} shows an example of the situation for one dimensional case.

\par
\smallskip
Let an integer vector ${\bm X}_i \in \mathbb{Z}^{n}_+$ 
denote the copy number of molecular species inside 
the voxel $\mathcal{V}_i~(i=1,2,\cdots,N)$. 
Stacking the vector ${\bm X}_i$, we define 
\begin{align}
{\bm X} := [{\bm X}_1^T, {\bm X}_2^T, \cdots,
{\bm X}_N^T]^T \in \mathbb{Z}^{nN}_+.
\end{align}
It is known that the stochastic time development of the molecular copy number
follows the chemical master equation (CME) for well-mixed chemical
systems \cite{Gillespie1992}.
Thus, the dynamics of chemical reactions inside the voxel $\mathcal{V}_i$ is 
given by 
%\begin{small}
\begin{align}
\frac{\partial {P}({\bm X}_i, t)}{\partial t}
\! = \!
%\sum_{i=1}^{N}
\sum_{r = 1}^{m_0}
&\left(
w_r({\bm X}_i \! -  \! 
{\bm s}_r) 
{P}({\bm X}_i  -  {\bm s}_r, t) \right.
%{P}({\bm X}  -  {\bm g}_i \otimes {\bm s}_r, t) \right. %\notag \\
% &
\notag  \\
&~~ \left. 
- 
w_r({\bm X}_i) {P}({\bm X}_i, t)
\right)
 = \mathcal{A}_i P({\bm X}_i, t),
\label{CME-eq}
\end{align}
%\end{small}
where $P({\bm X}_i, t)$ denotes the conditional probability that there are
${\bm X}_i$ molecules in $\mathcal{V}_i$ at time $t$ for a given initial state and time 
\footnote{More precisely, $P({\bm X}_i, t)$ should be written as $P({\bm X}_i, t | {\bm
X}_{i_0}, t_0)$ with the initial state ${\bm X}_{i_0}$ and time $t_0$. 
In this paper, however, we omit the condition, and simply 
write $P({\bm X}_i, t)$ to avoid notational complexity.}. 
The function $w_r(\cdot): \mathbb{Z}_+^n \rightarrow
\mathbb{R}_+~(r=1,2,\cdots,m_0)$ and 
the vector ${\bm s}_r \in
\mathbb{Z}^n$
denote the propensity function 
and stoichiometry for the reaction $\mathcal{R}_r$
 (see \cite{Iglesias2009} for details).
%$\mathcal{R}_r$ \cite{******}, and  denotes the 
%which represents the copy number of molecules changed by
%$\mathcal{R}_r$, and ${\bm g}_i := [0,\cdots,0,1,0,\cdots,0]^T \in \mathbb{Z}^N$ is a standard basis with 
%1 at the $i$-th entry. 
We define $\mathcal{A}_i$ as an infinitesimal generator describing the
development of $P({\bm X}_i, t)$.

% \begin{table}
% \centering
% \begin{tabular}{ l p{10cm} } \hline \hline
% $w_r(\cdot) \in \mathbb{Z}_+^n \rightarrow
% \mathbb{R}_+$ & propensity function for reaction
% $\mathcal{R}_r$ \\
% ${\bm s}_r \in \mathbb{Z}^n$ & 
% the stoichiometry for $\mathcal{R}_r$, 
% which represents the copy number of molecules changed by
% reaction $\mathcal{R}_r$. \\ \hline \hline
% \end{tabular}
% \end{table}

\par
\smallskip
The diffusion of molecules can be modeled by the exchange of molecules
between voxels (see Fig. \ref{scheme-fig}).
%Molecules diffuse from a voxel to an adjuscent voxel. 
When one molecule of $\mathcal{M}_k$ moves from the voxel $\mathcal{V}_i$ to $\mathcal{V}_{j}$, 
the number of molecules ${\bm X}$ changes to 
$[{\bm X}_1^T, \cdots, {\bm X}_i^T - {\bm e}_k^T, \cdots, {\bm X}_j^T +
{\bm e}_k^T, \cdots, {\bm X}_N^T]^T$, where 
${\bm e}_k:=[0,\cdots, 0, 1, 0,\cdots, 0] \in \mathbb{Z}^n$ is the
standard unit vector with 1 at the $k$-th entry.
Let $\mathcal{I}_i~(i=1,2,\cdots,N)$ denote a set of adjacent voxels 
of $\mathcal{V}_i$, {\it i.e.,} 
\begin{align}
\mathcal{I}_i := \{j \in \{1,2,\cdots,N\}~|~\mathcal{V}_j~\mathrm{is~adjacent~to~}\mathcal{V}_i.\}.
\end{align}
In general, ${\bm X}$ is updated as %by diffusion can be written as 
\begin{align}
{\bm X} + {\bm v}_{ij} \otimes {\bm e}_k, 
\end{align}
for each diffusion event, where the vector ${\bm v}_{ij} \in
\mathbb{Z}^{N}$ has $-1$ at the $i$-th entry, $+1$ at 
the $j$-th entries $(j \in \mathcal{I}_i)$ and 
0 at the other entries.
Thus, the dynamics of diffusion is written as  %is modeled as 
\begin{align}
\frac{\partial P({\bm X}, t)}{\partial t}
 = \sum_{i=1}^{N} \sum_{j \in \mathcal{I}_i} \sum_{k=1}^{n}
&\left(\delta_k (X_i \!+\! 1){P}({\bm X} \!-\! {\bm v}_{ij} \otimes {\bm e}_k, t) \right.  %\notag \\
% & 
\notag  \\
&\left. - \delta_k X_i {P}({\bm X},\! t)\right)
 \!\!=\! \mathcal{D}P({\bm X},\! t),
%\delta_{ij}^{(k)}({\bm X} - {\bm v}_{ij}^{(k)})
%\mathbb{P}({\bm X} - {\bm v}_{ij}^{(k)})
% - \delta_{ij}^{(k)}({\bm X}) \mathbb{P}({\bm X})
\label{diffusion-eq}
\end{align}
where $P{({\bm X}, t)}$ is the conditional probability that there are
${\bm X}$ molecules at time $t$ for a given initial state and time.
The diffusion constant of the molecule $\mathcal{M}_k$ is defined 
 by 
%\begin{align}
$\delta_k := {d_k}/{\ell^2}$
with a deterministic diffusion rate $d_k  > 0 $ and the characteristic
length of each voxel $\ell$. 
The probability that one molecule of $\mathcal{M}_k$ moves from
 $\mathcal{V}_i$ to an adjacent voxel within a small time interval $[t,
 t+\Delta t]$ is given by $\delta_k X_i \Delta t$.
We define $\mathcal{D}$ as the infinitesimal generator that accounts 
for diffusion.

\par
\smallskip
From (\ref{CME-eq}) and (\ref{diffusion-eq}), 
we have the following reaction-diffusion master equation (RDME). 
\begin{align}
\frac{\partial P({\bm X}, t)}{\partial t}
 = 
\sum_{i=1}^{N}\mathcal{A}_i P({\bm X}_i, t) + 
\mathcal{D} P({\bm X}, t).
\label{sys-eq}
\end{align}

% \medskip
% \noindent
% {\bf Remark 1.~}
% It is assumed in (\ref{sys-eq}) that molecules are 
% well-mixed in each voxel. 
% Thus, the volume $\Omega$ should be large enough to satisfy 
% this assumption.

\par
\smallskip
Although the RDME describes the dynamics of intrinsic noise in stochastic
reaction-diffusion systems, it is known that the RDME is hard to solve
analytically in most cases. 
Hence, in the next section, we introduce an approximation method to understand the properties of 
the intrinsic noise in an analytic way.
We hereafter restrict our attention to one dimensional spatial domain for
simplicity. The theoretical results, however, can be extended to higher
dimensional cases in a similar approach.

%%%%%%%%%%%%%%%%%%%%%%%%%%%%%%%%%%%%%%%%%%%%%%%%%%%%%%%%%%%%%%%%%%%%%%%%%%%%%%%%%%%%%%%%%%%%%%%%
\section{Linear noise approximation}
In this section, we briefly describe the idea of linear noise approximation
\cite{vanKampen2007}, and derive the approximated dynamics of (\ref{sys-eq}).
% Then, we present a dynamical model of the approximated
%intrinsic noise. % modeled by the RDME is 

\par
\smallskip
Let ${\bm x}^\Omega := {\bm X}/{\Omega} \in \mathbb{R}_+^{nN}$ denote the
concentration of molecules. 
Note that the molecular copy number ${\bm X}$  divided by the volume
$\Omega$ is the concentration.
It is known that ${\bm x}^\Omega$ converges to a deterministic solution 
in the thermodynamic limit $\Omega \rightarrow \infty$ \cite{Kurtz1971, Arnold1980}. 
Specifically, let 
\begin{align}
{\bm x}_i := \lim_{\Omega \rightarrow \infty} \frac{{\bm
 X}_i}{\Omega}~~(i=1,2,\cdots,N), 
\label{limit-eq}
\end{align}
and ${\bm x} := [{\bm x}_1^T, {\bm x}_2^T, \cdots, {\bm x}_N^T]^T \in
\mathbb{R}^{nN}_+$. 
Then, ${\bm x}(t)$ follows the spatially discretized
reaction-diffusion equation
\begin{align}
\dot{{\bm x}} = 
\begin{bmatrix}
f({\bm x}_1) \\
\vdots \\
f({\bm x}_N)
\end{bmatrix}
 + 
(L \otimes D)
{\bm x} 
\label{continuous-eq}
\end{align}
 with the initial value ${\bm x}(0):= \lim_{\Omega \rightarrow \infty}
 {\bm x}^\Omega(0)$, where 
$L$ is defined by a discretized Laplacian
 matrix 
\begin{align}
L := 
\begin{bmatrix}
-1 & 1 & 0 &\cdots & 0 \\
1 & -2 & 1 &\cdots & \vdots \\
\vdots & \ddots & \ddots & \ddots & \vdots \\
0 & \cdots & 1 & -2 & 1 \\
0 & \cdots & 0 & 1 & -1
\end{bmatrix}
 \in \mathbb{R}^{N \times N} 
\label{L-def}
\end{align}
and $D$ is the deterministic
diffusion rate matrix defined by 
\begin{align}
D := \mathrm{diag}[d_1, d_2, \cdots, d_n] \in \mathbb{R}^{n \times n}
\end{align}
with $d_i := \delta_i \ell^2$. % defined in (\ref{******}).
% It should be noted that the deterministic diffusion rate $d_i$ is different from 
% the diffusion constant $\delta_i$ in the stochastic model
% (\ref{sys-eq}). 
% The physical dimension of $d_i$ is $\mathrm{(length)^2}/(\mathrm{time})$, while 
% that of $\delta_i$ is $1/(\mathrm{time})$. 
% The detailed explanation can be found in chapter 9 of \cite{Edelstein-Keshet2005}.
In (\ref{L-def}), the Neumann boundary condition with zero flux is
assumed. Though the following argument can also be applied to the case where 
the boundary is periodic, we hereafter assume the Neumann boundary for 
simplicity.

\par
\smallskip
As we have seen in (\ref{limit-eq}), the variance of molecular copy
number ${\bm X}_i$ becomes less significant as the volume $\Omega$ becomes
larger, and eventually converges to zero.
Thus, one would expect from this observation that 
${\bm X}_i$ is approximated around the deterministic solution ${\bm x}_i$ as 
\begin{align}
{\bm X}_i \simeq \Omega {\bm x}_i + \Omega^{\frac{1}{2}} {\bm \eta}_i
~(i=1,2,\cdots,N), 
\label{simeq-eq}
\end{align}
where ${\bm \eta}_i$ is a noise term whose properties are specified
later.
It should be noted that $\Omega {\bm x}_i$ is dimensionless, 
and the noise grows with $O(\Omega^{{1}/{2}})$.
%Note also that the noise term converges to zero in the limit 
%$\Omega \rightarrow \infty$.

\par
\smallskip
Let ${\bm \eta} \in \mathbb{R}^{nN}$ be defined by 
\begin{align}
{\bm \eta} := 
[{\bm \eta}_1^T, {\bm \eta}_2^T, \cdots, {\bm \eta}_N^T]^T 
\in \mathbb{R}^{nN}. 
\end{align}
Note that the subscript of ${\bm \eta}$ stands for the index of 
a voxel.
Since ${\bm x}_i$ is a deterministic value obtained from 
 (\ref{continuous-eq}), 
${\bm \eta}$ is the only random variable that determines the 
stochastic fluctuations. 
Thus, defining $\Pi({\bm \eta}, t)$ as the probability 
distribution of ${\bm \eta}$ at time $t$, 
we have $P({\bm X}, t) \simeq \Omega^{\frac{-nN}{2}}\Pi({\bm \eta}, t)$, 
where the scaling term comes from the change of variable of 
probability distribution.
We see that the probability distribution $\Pi({\bm \eta}, t)$ 
contains the property of intrinsic noise.
%that is assumed to fluctuate around the deterministic solution $\Omega{\bm x}$.

\par
\smallskip
The idea of linear noise approximation \cite{vanKampen2007} is that 
we replace ${\bm X}$ in the equation (\ref{sys-eq}) with 
the right-hand side of (\ref{simeq-eq}), and 
expand around the deterministic value $\Omega {\bm x}$ 
with Taylor expansion.
Sorting with the power of $\Omega$, 
we have the deterministic model (\ref{continuous-eq}) 
in the term of $\Omega^{1/2}$, and a Fokker-Planck equation of 
$\Pi({\bm \eta}, t)$ in the term of $\Omega^{0}$.
Since $O(\Omega^{-1/2})$ terms become less significant as $\Omega$ 
becomes large, we can approximately adopt 
the Fokker-Planck equation as the dynamics of $\Pi({\bm \eta}, t)$. 
Then, the standard argument of stochastic systems % \cite{vanKampen2007} 
allows us to see that ${\bm \eta}$ satisfies the following 
dynamic equation.
\begin{align}
d{\bm \eta} = J_{{\bm x}(t)} {\bm \eta} dt + SW_{{\bm
 x}(t)}^{\frac{1}{2}} d\mathcal{B}(t),
\label{eta-eq}
\end{align}
where $\mathcal{B}(t)$ is a vector of 
$R := m_0 N + m_1 n$ independent standard Wiener process 
with $m_1 := \sum_{i=1}^{N} |\mathcal{I}_i|$. 
The matrix $J_{{\bm x}(t)} \in \mathbb{R}^{nN \times nN}$ is  
Jacobian of (\ref{continuous-eq}) at ${\bm x}(t)$.
$S$ and $W_{{\bm x}(t)}$ are defined as 
 stoichiometry matrix and a diagonal matrix with 
the propensity functions at diagonal entries, 
which are specified from (\ref{sys-eq}), respectively \cite{Iglesias2009}.

\par
\smallskip
In the following section, we consider the stochastic fluctuation 
around a spatially homogeneous equilibrium point, at which 
stochastic Turing pattern is expected. 
In particular, a characteristic structure of the stochastic system (\ref{eta-eq}) 
is revealed based on properties of the
reaction-diffusion system.
%, and its physical interpretation is presented.

\section{Analysis of noise-induced spatial patterns}
\subsection{Structure of the stochastic system and noise covariance}
\par
\smallskip
Let ${\bm x}_e \in \mathbb{R}^n$ denote an equilibrium of $\dot{\bm x}_i =
f({\bm x}_i)$, where $f(\cdot)$ is given in the deterministic model
defined in (\ref{continuous-eq}). 
We can see that 
${\bm x}^*:= [{\bm x}_e^T, {\bm x}_e^T, \cdots, {\bm x}_e^T]^T \in
\mathbb{R}_+^{nN}$ is a spatially homogeneous equilibrium point, {\it
i.e.}, the values are the same between voxels. %do not depend on the position. 
% {\color{red}Systems explanation first. and its interpretation as a
% multi-agent system. 
% Then, briefly say that $H_2$ norm corresponds to the noise covariance,
% and go to the main result.}
% In this section, we study the structure of the matrices in
% (\ref{eta-eq}) at the homogeneous equilibrium 
% in detail. 
% \medskip
% \noindent
% {\bf Remark 4.~} 
% Note that the deterministic system (\ref{continuous-eq}) always has a spatially
% homogeneous equilibrium point ${\bm x}^* := [{\bm x}_e^T, {\bm x}_e^T,
% \cdots, {\bm x}_e^T]^T$, where ${\bm x}_e \in \mathbb{R}^{n}$ satisfies 
% $f({\bm x}_e) = 0$. 

\begin{table}[tb]
\centering
\caption{Meaning of the matrices}
\label{meaning-table}
{\normalsize
\begin{tabular}{l p{11cm}} \hline \hline
$S_0$ & Stoichiometry matrix for $\mathcal{R}_1, \mathcal{R}_2, \cdots
 \mathcal{R}_{m_0}$ \\ \hline
$S_1$ & Incidence matrix of voxels \\ \hline
$W_0$ & Diagonal matrix whose entries are propensity functions for
		 $\mathcal{R}_1, \mathcal{R}_2, \cdots, \mathcal{R}_{m_0}$.\\ \hline
$W_1$ & Diagonal matrix whose entries are the equilibrium concentrations 
of $\mathcal{N}_1, \mathcal{N}_2, \cdots, \mathcal{N}_n$. \\ \hline
$D$ & Diagonal matrix whose entries are the diffusion rate of
		 $\mathcal{N}_1, \mathcal{N}_2, \cdots, \mathcal{N}_n$. \\ \hline \hline
\end{tabular}
}
\end{table}

\par
\smallskip
Using the structure of (\ref{sys-eq}), it can be shown that the matrices
in (\ref{eta-eq}) are given by 
\begin{align}
J_{{\bm x}^*} &:= I_N \otimes K + L \otimes D \in \mathbb{R}^{nN \times
 nN}, \notag \\
S &:= [I_N \otimes S_0, S_1 \otimes I_n] \in \mathbb{Z}^{nN \times R},
 \notag \\
W_{{\bm x}^*} &:= 
\begin{bmatrix}
I_N \otimes W_0 & O \\
O & I_{m_1} \otimes DW_{1}
\end{bmatrix}
\in
\mathbb{R}_+^{R \times R}, 
\notag
\end{align}
where 
\begin{align}
K &:= 
\left(
\frac{\partial f}{\partial {\bm x}}
\right)
\bigg|_{{\bm x} = {\bm x}^*}
\in \mathbb{R}^{n \times n}  \notag \\
S_0 &:= [{\bm s_1}, {\bm s_2}, \cdots, {\bm s}_{m_0}] \in \mathbb{Z}^{n
 \times m_0} \notag \\
S_1 &:= [{\bm v}_1, {\bm v}_2, \cdots, {\bm v}_{N}] \in \mathbb{Z}^{N
 \times m_{1}} \notag \\
W_0 &:= \mathrm{diag}(\hat{w}_1({\bm x}_e), \hat{w}_2 ({\bm x}_e),
 \cdots, \hat{w}_{m_{0}}({\bm x}_e)) \in \mathbb{R}^{m_0 \times m_0}
 \notag \\
W_1 &:= \mathrm{diag}(x_{e}^{(1)}, x_{e}^{(2)}, \cdots, x_{e}^{(n)}) \in\
 \mathbb{R}^{n \times n}. \notag
\end{align}
The vector ${\bm v}_i$ is ${\bm v}_i := [{\bm v}_{ij_1}, {\bm v}_{ij_2},
\cdots, {\bm v}_{ij_{|I_i|}}] \in \mathbb{Z}^{N \times |\mathcal{I}_i|}$ with $j_k \in \mathcal{I}_i$.
The function $\hat{w}_i(\cdot):\mathbb{R}_+^{n} \rightarrow
\mathbb{R}_+$ is defined as the deterministic reaction rate for 
reaction $\mathcal{R}_i$, and $x_e^{(i)}$ is 
the $i$-th entry of ${\bm x}_e~(i=1,2,\cdots,n)$. 
The meanings of each matrix are summarized in Table \ref{meaning-table}.

\begin{figure}
\centering
\includegraphics[clip, width=12cm]{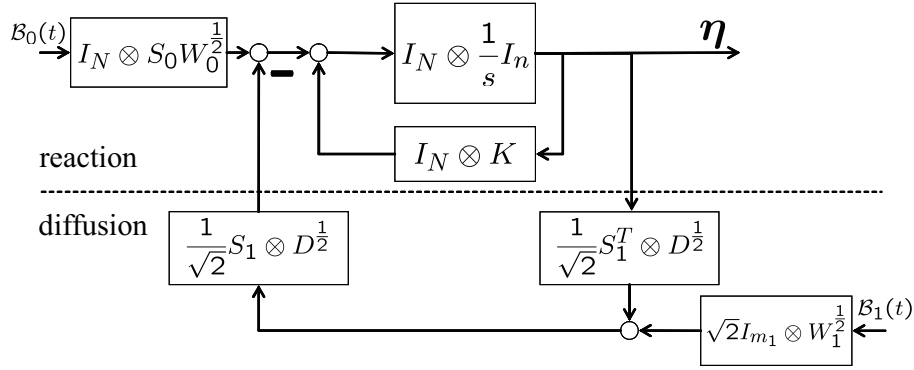}
\caption{Block diagram of the stochastic system (\ref{eta-eq}).}
\label{block-fig}
\end{figure}

\par
\smallskip
Substituting the above definition into (\ref{eta-eq}), 
we see that the overall system has the structure shown 
in Fig. \ref{block-fig}. 
Since the matrices $S_0, W_0$ and $K$ are associated with the reactions 
inside each voxel, $\mathcal{B}_0$ excites the intrinsic noise 
arising from intravoxel reactions. 
On the other hand, $\mathcal{B}_1$ excites the one arising from
diffusion (see Fig. \ref{block-fig}). 

\par
\smallskip
It follows from (\ref{eta-eq}) that 
the covariance of the noise $\mathbb{E}[{\bm \eta}{\bm \eta}^T]$ 
at the steady state ${\bm x}^*$ is given by the Lyapunov equation 
\begin{align}
J_{{\bm x}^*}\Sigma + \Sigma J_{{\bm x}^*}^T + SW_{{\bm x}^*}S^T = 0.
\end{align}
Thus, we have the following proposition. %the covariance of noise satisfies the the following proposition. 

\medskip
\noindent
{\bf Proposition 1.~}
{\it 
Consider the system (\ref{eta-eq}). 
Suppose the spatially homogeneous equilibrium ${\bm x}^*$ is 
locally stable. % in (\ref{continous-eq}). 
The steady state covariance of the noise $\Sigma := \mathbb{E}[{\bm \eta}{\bm
\eta}^T]$ is given by the positive definite solution of the
Lyapunov equation
\begin{align}
(I_N \otimes K + L \otimes D) \Sigma + \Sigma (I_N \otimes K^T + L
 \otimes D)
\notag \\
 + 
(
I_N \otimes S_0 W_0 S_0^T - 2L \otimes D W_1
)
=
0
\label{prop1-eq}
\end{align}
}

\medskip
\noindent
This proposition provides an approximation of the covariance of the intrinsic
noise whose dynamics is given by (\ref{eta-eq}).
The $i$-th $n$ by $n$ diagonal block of $\Sigma$ 
stands for the covariance inside the voxel $\mathcal{V}_i$, 
{\it i.e.,} $\mathbb{E}[{\bm \eta}_i {\bm \eta}_i^T]$.
The covariance between voxels appears in off-diagonal entries, 
where the $(i,j)$ off-diagonal block is defined by $\mathbb{E}[{\bm \eta}_i
{\bm \eta}_j]$.
It should be noted that $\mathrm{Tr}(\Sigma)$ in Proposition 1 provides
$\mathcal{H}_2$ norm of the system in Fig. \ref{block-fig}.

\medskip
\noindent
{\bf Remark 1.~}
It is interesting to observe the system in Fig. \ref{block-fig} 
from a viewpoint of dynamical multi-agent systems.
In Fig. \ref{block-fig}, the blocks in the reaction part (upper blocks) are
homogeneous block diagonal matrices, and each block diagonal 
entry $H(s):=(sI-K)^{-1}$ corresponds to the linearized dynamics of 
%the reactions in each voxel. 
intravoxel reactions.
It is clear that $\mathcal{B}_0$ is the noise that perturbs the states of
each subsystem $H(s)$.
The blocks in the diffusion part (lower blocks), on the other hand, describes the structure
of information exchange between $H(s)$. 
Since $S_1$ is the incidence matrix 
associated with 
%corresponding to 
the graph Laplacian $L$, 
$\mathcal{B}_1$ can be considered as a perturbation to the sensed
output ${\bm \eta}_{i\!-\!1}\! -\! {\bm \eta}_i$. 
Thus, one might expect that the study of this class of system is useful 
for engineering applications as well. 
$\hfill \Box$

%%%%%%%%%%%%%%%%%%%%%%%%%%%%%%%%%%%%%%%%%%%%%%%%%%%%%%%%%%%%%%%%%%%%%%%%%%%%%%%%
\subsection{Spatial power spectrum analysis of the intrinsic noise}
Although the matrix $\Sigma$ in (\ref{prop1-eq}) contains all information about the spatial 
covariance of ${\bm \eta}$, it is not easy to see 
the existence and profiles of spatial pattern at a glance. 
Hence, we here consider spatial spectrum analysis based on Proposition 1.

% Spatial patterns of the noise can be captured from 
% It is convenient to analyze spatial frequency of the noise 
% to capture the spatial pattern of the noise can be 
% In this section, we show that spatial power spectrum of the noise 
% can be computed from relatively small dimensional Lyapunov
% equations****. 

%\subsection{Spatial frequency analysis}

\begin{figure}
\centering
\includegraphics[clip,width=12cm]{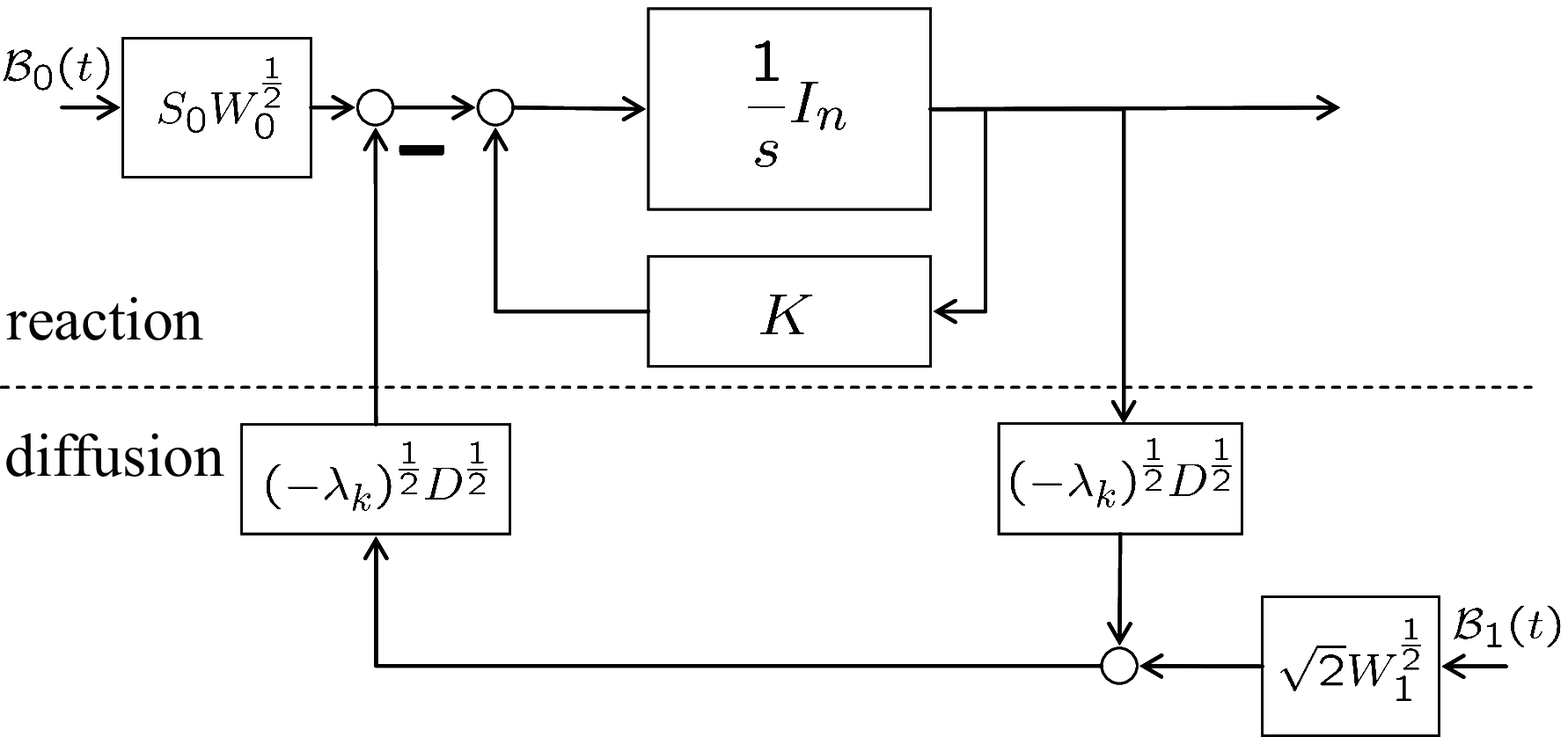}
\caption{The decomposed subsystem obtained from the large system in
 Fig. \ref{block-fig}.}
\label{block-fig2}
\end{figure}

\par
\smallskip
The following theorem presents an efficient computation method of the
spatial power spectrum of the noise. 

\medskip
\noindent
{\bf Theorem 1.~}
{\it 
Consider the system (\ref{eta-eq}). 
Suppose the spatially homogeneous equilibrium ${\bm x}^*$ is 
locally stable. % in (\ref{continous-eq}). 
Let ${\bm \xi}_i~(i=1,2,\cdots, N)$ denote the spatial frequency 
components of the steady state noise ${\bm \eta}_i$, {\it i.e.}
\begin{align}
{\bm \eta}_i
 \!=\! 
\sqrt{\frac{1}{N}}{\bm \xi}_1
 \!+\! 
\sqrt{\frac{2}{N}}\sum_{k=2}^{N}
\cos
\left( \frac{(k\!-\!1)\pi}{N} \left( i \!-\! \frac{1}{2}
\right) \right) {\bm \xi}_k. 
\end{align}
Then, the power spectral density $\Xi_k := \mathbb{E}[{\bm \xi}_k {\bm
\xi}_k^T] \in \mathbb{R}^{n \times n}$ at the frequency $\omega_k :=
(k-1)\pi/N~(k=1,2,\cdots,N)$ is given by the positive definite
solution of 
%{\small
\begin{align}
(K + \lambda_k D)\Xi_k + \Xi_k(K^T + \lambda_k D) 
+ S_0 W_0 S_0^T \notag \\
- 2  \lambda_k D W_1 =  0,
\label{theo1-eq}
\end{align}
where
\begin{align}
\lambda_k := -4 \sin^2 \left(\frac{(k-1)\pi}{2N}\right).
\end{align}
%}%\small
}%\it

% \medskip
% \noindent
% {\bf Proof.~}
% Let $U \in \mathbb{R}^{N \times N}$ be defined by 
% \begin{align}
% (U)_{ij} := 
% \begin{cases}
% \displaystyle
%  \sqrt{\frac{1}{N}} & (j=1) \\
% \displaystyle
%  \sqrt{\frac{2}{N}} \cos 
% \left(
% \frac{(k-1)\pi}{N} \left( i - \frac{1}{2} \right)
% \right) & 
% \mathrm{(otherwise)}
% \end{cases}
% ,
% \notag
% \end{align}
% where $(U)_{ij}$ stands for the $(i,j)$-th entry of $U$.
% We can see that the columns of $U$ are the eigenvectors of $L$. 
% Thus, we have 
% \begin{align}
% (I_N \otimes K + \Lambda \otimes D) \Xi + \Xi (I_N \otimes K^T + \Lambda
%  \otimes D) \notag \\
% + (I_N \otimes S_0 W_0 S_0^T - 2 \Lambda \otimes D W_1) = 0 
% \label{large-lyap-eq}
% \end{align} 
% by multiplying $U^T \otimes I_n$ and $U \otimes I_n$ from the left and 
% right of (\ref{large-lyap-eq}), respectively,  
% where $\Lambda := \mathrm{diag}(\lambda_1, \lambda_2, \cdots,
% \lambda_N)$ and $\Xi := (U^T \otimes I_n) \Sigma (U \otimes I_n)$.
% Note that we used $(A \otimes B)(C \otimes D) = AC \otimes BD$
% \cite{Bernstein2005} in the derivation.
% We see that Lyapunov solution $\Xi$ becomes a block diagonal matrix 
% $\Xi = \mathrm{blockdiag}(\Xi_1, \Xi_2, \cdots, \Xi_N)$, 
% because both $D$ and $W_1$ are diagonal. 
% $\hfill \Box$

\medskip
%The proof can be found in \cite{CDC2012Tech}.
We see from Theorem 1 that the spatial power spectral density of 
the molecule $\mathcal{M}_i$ is obtained as the $(i,i)$-th entry 
of the matrices $\Xi_k~(k=1,2,\cdots,N)$. 
%It is clear from the proof that 
In fact, $\Xi$ is obtained by applying discrete cosine transform 
to the covariance matrix $\Sigma$.
%, which can be corroborated from Wiener-Khinchin theorem as well. 
It should be noted that $\Xi_k~(k=1,2,\cdots,N)$ are 
obtained by solving $N$ Lyapunov equations of the size $n$ by $n$, 
which is more computationally efficient than solving $nN$ by $nN$ Lyapunov equation
(\ref{prop1-eq}) and applying discrete cosine transform.

\par
\smallskip
From a control theoretic viewpoint, the transformation presented
in Theorem 1 corresponds to the decomposition of the system 
into $N$ subsystems as shown in Fig. \ref{block-fig2}, where
$\lambda = \lambda_1, \lambda_2, \cdots, \lambda_N$. 
We see that the feedback gain $\lambda$ corresponds to 
frequency $\omega_k~(k=1,2,\cdots,N)$, and the spatial spectral density $\Xi_k$ varies in
terms of $\lambda$.
Thus, the computation of the spatial power spectrum is essentially the
same as $\mathcal{H}_2$ norm computation for various feedback gains $\lambda$.

\par
\smallskip
% The above observation allows us to decompose the Lyapunov equation
% (\ref{******}) into the sum of two Lyapunov solutions.
Since the noise $\mathcal{B}_0$ and $\mathcal{B}_1$ are decoupled in
Fig. \ref{block-fig2}, the $\mathcal{H}_2$ norm can be independently
computed for each input. 
Thus, we have the following proposition.

\medskip
\noindent
{\bf Proposition 2.~}
{\it 
The Lyapunov solution $\Xi_k$ in (\ref{theo1-eq}) can be 
decomposed into $\Xi_k = \Xi_{k1} + \Xi_{k2}~(k=1,2,\cdots,N)$, 
where $\Xi_{k1}$ and $\Xi_{k2}$ are the positive definite solutions 
of 
\begin{align}
(K + \lambda_k D) \Xi_{k1} + \Xi_{k1} (K^T + \lambda_k D) + S_0 W_0
 S_0^T =  0, \label{prop2-eq1}\\
(K + \lambda_k D) \Xi_{k2} + \Xi_{k2} (K^T + \lambda_k D) - 2\lambda_k
 DW_1 = 0.
\label{prop2-eq2}
\end{align}
}

\noindent
It should be noted that $\Xi_{k1}$ and $\Xi_{k2}$ represent the spectral density
of the noise originating from 
intravoxel reactions and
%the reactions inside the voxels and 
diffusion, respectively. 
Thus, Proposition 2 implies that we can independently analyze the contribution of
these noises.

\medskip
\noindent
{\bf Remark 2.~}
Theorem 1 and Proposition 2 can be applied for systems with the periodic
boundary condition. 
In that case, the Laplacian eigenvalues should be alternatively defined 
as $\lambda_k = -4 \sin^2((k-1)\pi/N)$, and the solution $\Xi$ 
corresponds to the discrete Fourier transform of $\Sigma$.
$\hfill \Box$

\section{Numerical Simulations}
In this section, we first illustrate stochastic Turing patterns induced
by intrinsic noise with the well-known Gray-Scott model \cite{Gray1984, Pearson1993}.
Then, we show that the spatial power spectrum analysis presented in the
previous section can capture the noisy spatial patterns. 
Furthermore, we present underlying mechanism of 
noise-induced Turing patterns based on the system theoretic
interpretation given in the previous sections.
Due to the limitation of the space, some details are left for 
our future publication. %omitted in this paper. 
%More detailed descriptions can be found in our future publication.%\cite{CDC2012Tech}.

%  that 
% the system theoretic interpretation 
% provides an intuitive explanation of the numerical simulation result.
% using the system
% we also present 
% Furthermore, we consider the mechanism***** of the pattern formation 
% from a control theoretic viewpoint. 

\begin{table}[tb]
\centering
\caption{Reactions of the Gray-Scott model}
\label{reaction-table}
{\normalsize
\begin{tabular}{|c|c|c|} \hline
 & Reaction & Propensity $w_i(\cdot)$ \\ \hline
$\mathcal{R}_1$ & $U_i + 2V_i \rightarrow 3V_i$ & $ k_1 [U_i] [V_i] ([V_i] - 1)/2\Omega^2$ \\
$\mathcal{R}_2$ & $V_i \rightarrow P_i$ & $k_2 [V_i]$ \\
$\mathcal{R}_3$ & $\phi \rightarrow U_i$ & $k_a u_0 \Omega$ \\
$\mathcal{R}_4$ & $U_i \rightarrow \phi$ & $k_a  [U_i]$ \\ 
$\mathcal{R}_5$ & $V_i \rightarrow \phi$ & $k_a [V_i]$ \\
\hline
 & $U_{i} \rightarrow U_{j}~~(\mathrm{if}~j \in \mathcal{I}_i)$ & $d_1/\ell^2$ \\
 & $V_{i} \rightarrow V_{j}~~(\mathrm{if}~j \in \mathcal{I}_i) $ & $d_2/\ell^2$ \\ \hline
\end{tabular}
}
\end{table}

\subsection{Noise-induced Turing patterns with the Gray-Scott model}
% The Gray-Scott model was originally presented in \cite{Gray1984} 
% as a pair of ordinary differential equations that describes 
% the dynamics of autocatalytic chemical reactions of two species 
% $U$ and $V$ in a spatially
% homogeneous, or well-mixed, environment.
% Later, Pearson \cite{Pearson1993} numerically demonstrated that the 
% same reaction scheme can present complex spatial patterns 
% when there is an additional diffusion term.

%based  on a reaction-diffusion model.

\par
\smallskip
The Gray-Scott model 
consists of the five chemical reactions in Table \ref{reaction-table}.
In reaction $\mathcal{R}_1$, ${U}$ is converted into
${V}$, where ${V}$ catalyzes the production of itself.
Then, the reaction $\mathcal{R}_2$ changes ${V}$ into the
final product form ${P}$. 
The reactions $\mathcal{R}_3, \mathcal{R}_4$ and $\mathcal{R}_5$
describe the inflow of ${U}$ from outside, degradation of
${U}$ and degradation of ${V}$.

% \par
% \smallskip
% The propensity functions for these reactions are shown in the right
% column of Table \ref{reaction-table}, where the subscript $i$ specifies the index
% of a voxel, and $[U_i]$ and $[V_i]$ denote the molecular copy number of
% $U_i$ and $V_i$. 
% The constants $u_0, k_a, k_1, k_2, d_1$ and $d_2$ are deterministic
% reaction rates, of which the details are not described here.

\par
\smallskip
The deterministic reaction-diffusion model of the Gray-Scott model is 
given by 
\begin{align}
\begin{array}{ll}
\displaystyle
\frac{\partial u}{\partial t}
 &= -uv^2 + a(1 - u) + d \nabla^2 u \vspace{2mm}\\
\displaystyle
\frac{\partial v}{\partial t} &= uv^2 - (a + k)v + \nabla^2
 v, 
\end{array}
\label{ai-eq}
\end{align}
% Let $u \in \mathbb{R}_+$ and $v \in \mathbb{R}_+$ denote 
% normalized concentrations of $\mathcal{U}$ and $\mathcal{V}$.
% The well-known deterministic Gray-Scott model is given by 
where $u \in \mathbb{R}_+$ and $v \in \mathbb{R}_+$ are 
normalized concentrations of $U$ and $V$. 
The dimensionless constants $a, k$ and $d$ are normalized reaction rates
and a diffusion rate defined %by
with the reaction constants in Table \ref{reaction-table} as 
\begin{align}
a := \frac{k_a}{k_1 u_0^2}, b:= \frac{k_2}{k_1 u_0^2}, 
d := \frac{d_1}{d_2}.
\end{align}

% \par
% \smallskip
% It is known that the deterministic system has multiple homogeneous equilibrium when 
% $z := 1 - 4(a + b)^2/a \ge 0$ \cite{Mazin1996}.
% Specifically, the equilibrium points are 
% \begin{align}
% (u_\star, v_\star) &= (1,0),  \notag \\
% (u_+, v_+) &= \left(\frac{1}{2}\left(1 - \sqrt{z}\right), \frac{a}{2(a + b)}
%  \left( 1 + \sqrt{z}\right)\right), \notag \\
% (u_-, v_-) &= \left(\frac{1}{2}\left(1 +  \sqrt{z}\right), \frac{a}{2(a + b)}
%  \left( 1 - \sqrt{z}\right)\right). \notag 
% \end{align}
% It was shown that $(u_\star, v_\star)$ and $(u_-, v_-)$ are always
% stable and unstable, respectively. The stability of $(u_+, v_+)$,
% however, depends on parameters. 
% Hence, $(u_+, v_+)$ can present Turing bifurcation.

% \par
% \smallskip
% Using linear stability analysis, the parameter region for 
% the existence of Turing pattern  was extensively studied in
% \cite{Mazin1996}. 
% Figure \ref{stability_region_fig} shows the parameter region for Turing instability 
% in terms of $a$ and $b$, where $d$ is set as $d = 6.0$.
% Note that the Turing instability condition is equivalent to the
% condition that $K$ is Hurwitz, and  
% \begin{align}
% K + \lambda D = 
% \begin{bmatrix}
% -a - v_*^2 + \lambda d & -2u_* v_*\\
% v_*^2 & -(a+b) + 2u_*v_* + \lambda
% \end{bmatrix}
% \notag
% \end{align}
% has a real pole in the open right half plane for some $\lambda < 0$. 

%where $K$ and $D$ are defined in (\ref{*****}).

\par
\begin{figure}
\centering
\includegraphics[clip,height=5.0cm]{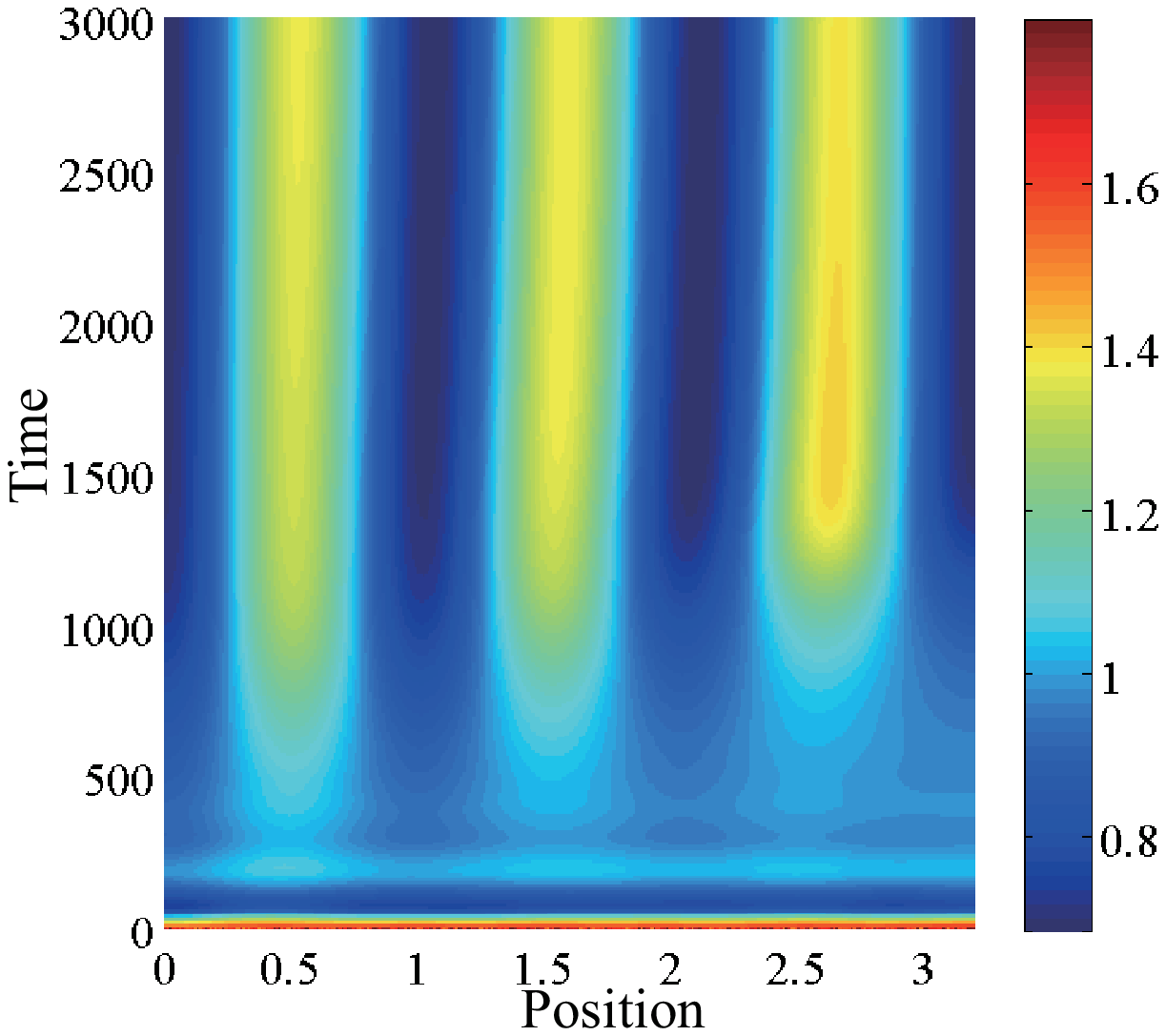}
\includegraphics[clip,height=5.0cm]{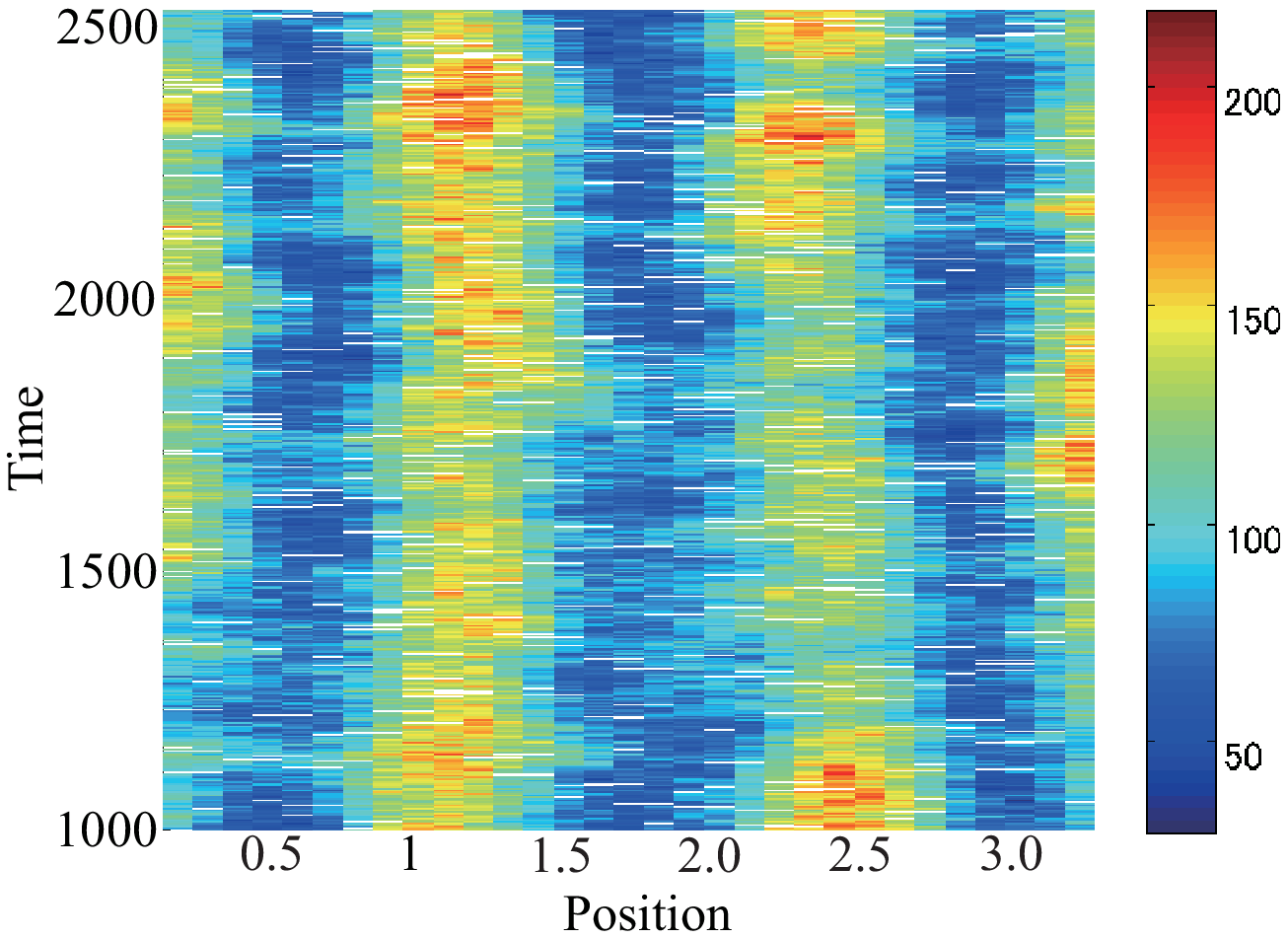}
\caption{Time development of the concentrations/copy numbers of 
$U$ for the first parameter set. (Left) A simulation result of the deterministic model. 
(Right) A result of the stochastic simulation.}
\label{param1-fig}
\end{figure}

\smallskip
Let the parameters be set as $(u_0, k_1, k_2, k_a, d_1, d_2) = (3.0, 4.0 \times 10^{-2}, 1.98 \times
10^{-2}, 2.16 \times 10^{-2}, 2.16 \times 10^{-13}, 3.6 \times
10^{-14})$, which corresponds to $(a,b,d) = (6.0 \times 10^{-2}, 5.5
\times 10^{-2} ,6.0)$. 
% In \cite{*****}, the parameter region for the existence
% of Turing pattern was extensively studied by linear stability analysis, 
% and we see that the system admits spatial patterns with the above
% parameters.
With this parameter set, both deterministic and stochastic simulations exhibit
the spatial pattern as shown in Fig. \ref{param1-fig}. 
For the stochastic simulation, the Gillespie's algorithm \cite{Gillespie1977}
was used, where the domain was partitioned into $N = 32$
voxels, and the characteristic length and the volume were $\ell =
1.0 \times 10^{-1}$ and $\Omega = 1.0 \times 10^{2}$, respectively.

\par
\begin{figure}
\centering
\includegraphics[clip,height=5.0cm]{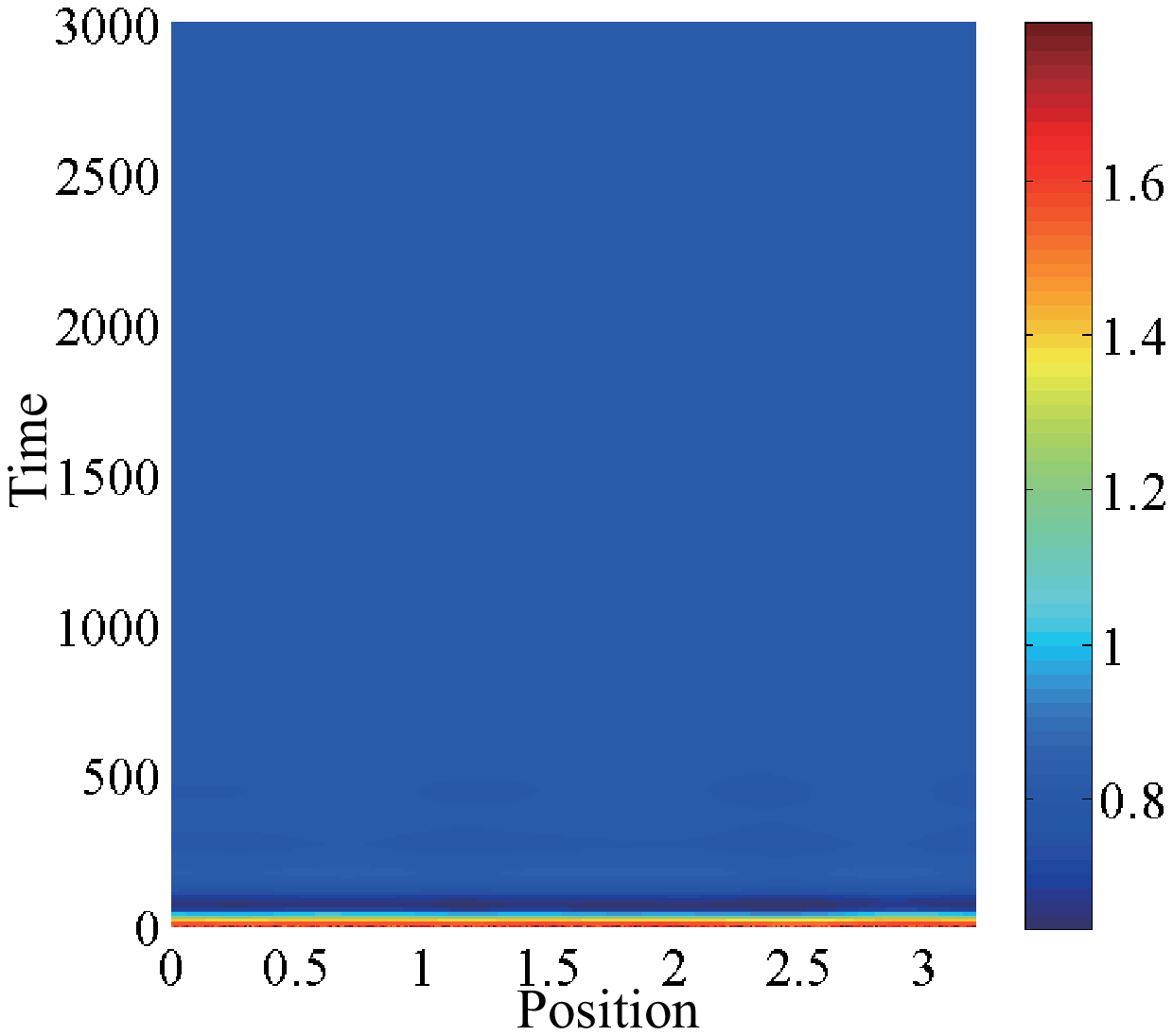}
\includegraphics[clip,height=5.0cm]{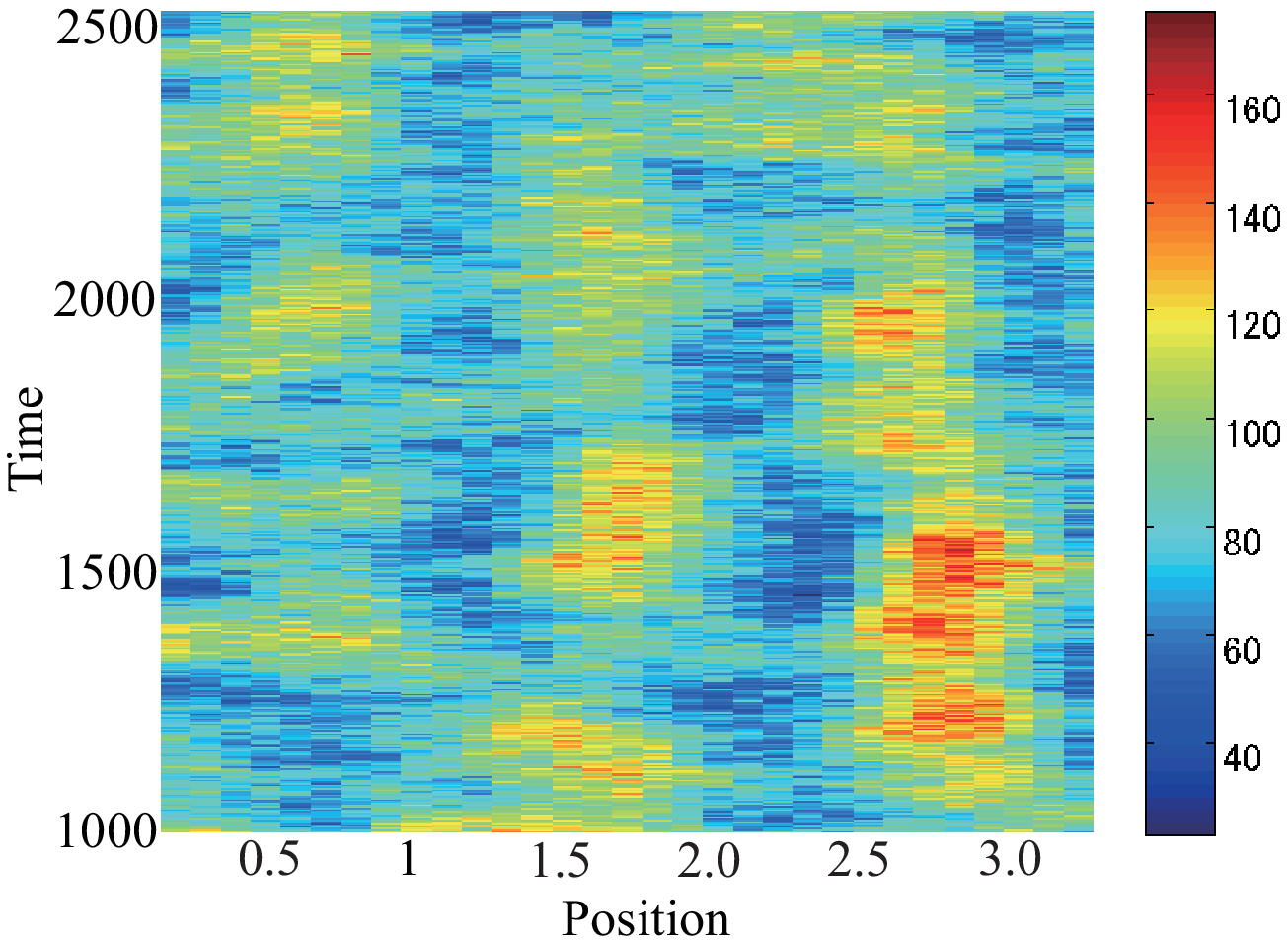}
\caption{Time development of the concentrations/copy numbers of 
$U$ for the second parameter set. (Left) A simulation result of the deterministic model. 
(Right) A result of the stochastic simulation. The red and blue patches
 imply the existence of noise-induced spatial pattern.}
\label{param2-fig}
\end{figure}

\smallskip
We now change the parameter $k_2$ in the above example to $1.76 \times
10^{-2}$, which results in $(a,b,d) = (6.0 \times 10^{-2}, 4.9 \times 10^{-2},
6.0)$. According to the deterministic analysis \cite{Mazin1996}, a 
spatially homogeneous equilibrium is stable, hence no spatial pattern is
expected.
In fact, the deterministic model converges to a homogeneous equilibrium
(Fig. \ref{param2-fig} (Left)). 
However, we observe that the stochastic simulation in Fig. \ref{param2-fig} (Right) still displays the spatial pattern.
%On the other hand, the stochastic simulation result shown in
%Fig. \ref{******} exhibits noisy but spatially inhomogeneous steady
%state. 
It should be emphasized that the deterministic model (\ref{ai-eq}) failed to capture this
spatial pattern.

\begin{figure}
\centering
\includegraphics[clip,width=6cm]{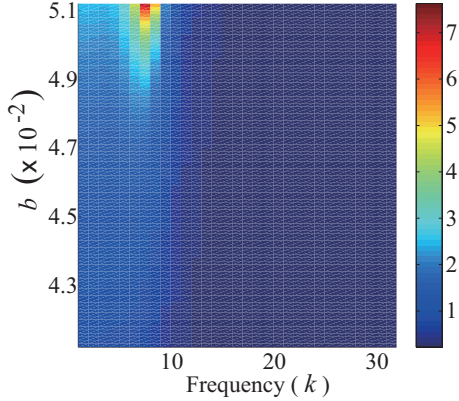}
\caption{Spatial power spectrum of intrinsic noise ${\bm \eta}$
 computed from Theorem 1. The horizontal axis is $k$ in $\omega_k =
 (k-1)\pi/N$.}
\label{frequency-fig}
\end{figure}

\begin{figure}[tb]
\centering
\includegraphics[clip,width=7cm]{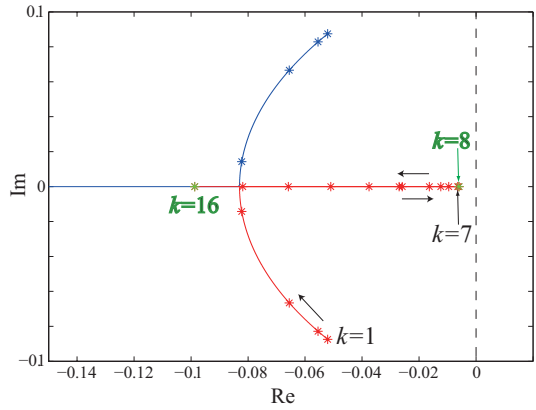}
\caption{Trajectory of the eigenvalues of $K + \lambda D$ in terms of
 $\lambda$. The mark $*$ stands for the eigenvalues at $\lambda =
 \lambda_k$.}
\label{eigenvalue-fig}
\end{figure}

% \begin{figure*}
% \centering
% \includegraphics[clip,width=4.8cm]{./figure/frequency_all.eps}
% \includegraphics[clip,width=4.8cm]{./figure/frequency_reaction.eps}
% \includegraphics[clip,width=4.8cm]{./figure/frequency_diffusion.eps}
% \caption{Spatial power spectrum of intrinsic noise. The horizontal axis 
% is $k$ in $\omega_k = (k-1)\pi/N$.
% (Left) Power
%  spectrum of ${\bm \eta}$ obtained from Theorem 1. (Center) 
% Power spectrum of the intrinsic noise arising from intravoxel reactions
%  (${\Xi}_{k_1}$ in Proposition 2). (Right) Power spectrum of 
% the intrinsic noise arising from diffusion (${\Xi}_{k_2}$ in
%  Proposition 2)}
% \label{frequency-fig}
% \end{figure*}

% \begin{figure}
% \centering
% \includegraphics[clip,
%  width=6cm]{./figure/turing_instability_region.eps}
% \caption{Turing instability region for the Gray-Scott model obtained 
% in \cite{Mazin1996}}
% \label{stability_region_fig}
% \end{figure}

\subsection{Spatial power spectrum analysis}

In this section, we first show that Theorem 1 and Proposition 2 
can capture the stochastic spatial patterns 
shown in Fig. \ref{param2-fig} (Right), then illustrate the mechanism of the stochastic pattern generation 
from a system theoretic point of view.

\par
\smallskip
Consider the case where the deterministic system (\ref{ai-eq}) does
not present Turing pattern. 
Using Theorem 1, we can approximately compute the steady state spatial power spectrum.
Figure \ref{frequency-fig} shows the computation result in terms of $b$.
% compute the solution of (\ref{*****}) in Theorem 1, 
% see how  changes for various parameters. 
%The result is shown in Fig. \ref{frequency-fig} (Left). 
We see that as $b$ becomes larger, 
sharp peak appears at $k=6$ and $7$, which correspond to 
$\omega_6 = 5\pi/32$ and $\omega_7 = 6\pi/32$. 
For $b = 4.9 \times 10^{-2}$, which is the second parameter set 
in the previous example, the peak frequency is $\omega_7$.  
Hence, we can expect spatial patterns with three periods.
Indeed, the spatial structure can be confirmed in Fig. \ref{param2-fig} (Right).
The ingredient of the power spectrum can be analyzed with Proposition 2,
and it can be concluded that both reactions and diffusion contribute to 
form the peak at non-zero frequency in Fig. \ref{frequency-fig} %(see \cite{CDC2012Tech} for the details). 

\par
\smallskip
It is interesting to consider these observations from a 
control theoretic viewpoint.
%As we have seen in the previous section, 
In general, the deterministic Turing instability is determined from the stability of 
$K + \lambda D$, which corresponds to 
the stability of the system in Fig. \ref{block-fig2}.
Figure \ref{eigenvalue-fig} illustrates the eigenvalue distribution 
of $K+\lambda D$ in terms of $\lambda$. 
Note that this corresponds to drawing 
the poles of the system in Fig. \ref{block-fig2} by changing 
the feedback gain $\lambda$.
We see that one eigenvalue approaches to the imaginary axis 
as $k$ increases, and the largest real part is achieved at $k=8$ 
(see Fig. \ref{eigenvalue-fig}). 
In view of the stability of this system, 
it is natural that the largest $\mathcal{H}_2$ norm 
is obtained around $k=8$. 
This observation corroborates the aforementioned analysis 
illustrated in Fig. \ref{frequency-fig}, where we actually have the peak at 
$k=7$. 
%Further consideration can be found in \cite{CDC2012Tech}.

%\par
%\smallskip
% In addition, as we increase $b$, the system tends to become 
% unstable as shown in Fig. \ref{stability_region_fig}. 
% Thus, it is expected that the largest real part of the 
% eigenvalue in Fig. \ref{eigenvalue-fig} comes close to the imaginary axis.  
% Consequently, $\mathcal{H}_2$ norm for the corresponding 
% $\lambda (= \lambda_k)$ could present a distinctively a 
% large value. 
% This peak corresponds to the red area around $b = 5.1$ 
% in Fig. \ref{frequency-fig} (Left).

% \begin{align}
% S_0 W_0 S_0 - 2\lambda_k DW_1= 
% \begin{bmatrix}
% k_1 u_+ v_+^2 + k_a u_0 + k_a u_+ - 2\lambda_k d_1 u_+ & -k_1 u_+ v_+^2 \\
% -k_1 u_+ v_+^2 & k_1 u_+ v_+^2 + k_2 v_+ + k_a v_+ -2 \lambda_k d_2 v_+
% \end{bmatrix}
% \end{align}

\section{Conclusion}
We have analyzed
noise-induced pattern formation under 
the reaction-diffusion kinetics described by the RDME.
%In particular, 
%intrinsic noise under reaction-diffusion kinetics. 
Using the linear noise approximation, 
we have revealed the characteristic structure of 
the stochastic reaction-diffusion system, and 
presented the underlying mathematical structure 
of noise-induced pattern formation from a control theoretic point of view. 
The analytic tool to examine the spatial power spectrum 
of intrinsic noise  has also been derived in the analysis, and its 
effectiveness has been confirmed through the numerical simulations.

% together with the spatial frequency analysis
% and its connection with pattern formation has been explained 
% Then, 
% it is shown that the computation of the covariance of intrinsic
% noise can be viewed as the $\mathcal{H}_2$ norm computation of a coupled 
% dynamic multi-agent system, where disturbance inputs are injected to 
% each agent's states and sensed inputs.
% Finally, Gray-Scott****
% Although the result is a speacial case.
%The same arguemnt can be applied to two-species model..

\smallskip
\noindent
{\bf Acknowledgments: }
This work was supported in part by the Ministry of Education, Culture, Sports, Science and Technology in
Japan through Grant-in-Aid for Scientific Research (A) No. 21246067 and Grant-in-Aid for JSPS Fellows No. 23-9203.
% This work is supported in part by Grant-in-Aid for JSPS Fellows
%  of Japan Society for the Promotion of Science (JSPS) under grant
% No. 23-9203. 

%%%%%%%%%%%%%%%%%%%%%%%%%%%%%%%%%%%%%%%%%%%%%%%%%%%%%%%%%
%%%%%%%%%%%%%%%%%%%%%%%%%%%%%%%%%%%%%%%%%%%%%%%%%%%%%%%%%
%% Bibliography
%%%%%%%%%%%%%%%%%%%%%%%%%%%%%%%%%%%%%%%%%%%%%%%%%%%%%%%%%
%%%%%%%%%%%%%%%%%%%%%%%%%%%%%%%%%%%%%%%%%%%%%%%%%%%%%%%%%
\bibliographystyle{ieeetran}
%{\color{red}
\bibliography{20120307.bib}

% Generated by IEEEtran.bst, version: 1.13 (2008/09/30)
\begin{thebibliography}{10}
\providecommand{\url}[1]{#1}
\csname url@samestyle\endcsname
\providecommand{\newblock}{\relax}
\providecommand{\bibinfo}[2]{#2}
\providecommand{\BIBentrySTDinterwordspacing}{\spaceskip=0pt\relax}
\providecommand{\BIBentryALTinterwordstretchfactor}{4}
\providecommand{\BIBentryALTinterwordspacing}{\spaceskip=\fontdimen2\font plus
\BIBentryALTinterwordstretchfactor\fontdimen3\font minus
  \fontdimen4\font\relax}
\providecommand{\BIBforeignlanguage}[2]{{%
\expandafter\ifx\csname l@#1\endcsname\relax
\typeout{** WARNING: IEEEtran.bst: No hyphenation pattern has been}%
\typeout{** loaded for the language `#1'. Using the pattern for}%
\typeout{** the default language instead.}%
\else
\language=\csname l@#1\endcsname
\fi
#2}}
\providecommand{\BIBdecl}{\relax}
\BIBdecl

\bibitem{Turing1952}
A.~M. Turing, ``The chemical basis of morphogenesis,'' \emph{Philosophicals
  Transactions of the Royal Society of London B}, vol. 237, no. 641, pp.
  37--72, 1952.

\bibitem{Kondo2010}
S.~Kondo and T.~Miura, ``Reaction-diffusion model as a framework for
  understanding biological pattern formation,'' \emph{Science}, vol. 329, pp.
  1616--1620, 2010.

\bibitem{Elowitz2002}
M.~B. Elowitz, A.~J. Levine, E.~D. Siggia, and P.~A. Swain, ``Stochastic gene
  expression in a single cell,'' \emph{Nature}, vol. 297, no. 5584, pp.
  1183--1186, 2002.

\bibitem{Gillespie1992}
D.~T. Gillespie, ``A rigorous derivation of the chemical master equation,''
  \emph{Physica A}, vol. 188, no. 1--3, pp. 404--425, 1992.

\bibitem{Rudner2010}
D.~Z. Rudner and R.~Losick, ``Protein subcellular localization in bacteria,''
  \emph{Cold Spring Harbor Perspectives in Biology}, vol.~2, no. a000307, 2010.

\bibitem{Gardiner1976}
C.~W. Gardiner, K.~J. McNeil, D.~F. Walls, and F.~S. Matheson, ``Correlations
  in stochastic theories of chemical reactions,'' \emph{Journal of Statistical
  Physics}, vol.~14, no.~4, pp. 307--331, 1976.

\bibitem{Biancalani2010}
T.~Biancalani, D.~Fanelli, and F.~D. Patti, ``Stochastic {Turing} patterns in
  the {Brusselator} model,'' \emph{Physical Review E}, vol.~81, no. 046215,
  2010.

\bibitem{Glandsdorff1971}
P.~Glandsdorff and I.~Prigogine, \emph{Thermodynamic theory of structure,
  stability and fluctuations}.\hskip 1em plus 0.5em minus 0.4em\relax Wiley,
  New York, 1971.

\bibitem{Butler2011}
T.~Butler and N.~Goldenfeld, ``Fluctuation-driven {Turing} patterns,''
  \emph{Physical Review E}, vol.~84, no. 011112, 2011.

\bibitem{Biancalani2011}
T.~Biancalani, T.~Galla, and A.~J. McKane, ``Stochastic waves in a
  {Brusselator} model with nonlocal interaction,'' \emph{Physical Review E},
  vol.~84, no. 026201, 2011.

\bibitem{vanKampen2007}
N.~G. van Kampen, \emph{Stochastic processes in physics and chemistry},
  3rd~ed.\hskip 1em plus 0.5em minus 0.4em\relax North Holland, 2007.

\bibitem{Scott2011}
M.~Scott, F.~J. Poulin, and H.~Tang, ``Approximating intrinsic noise in
  continuous multispecies models,'' \emph{Proceedings of the Royal Society A},
  vol. 467, no. 2127, pp. 718--737, 2011.

\bibitem{Gray1984}
P.~Gray and S.~K. Scott, ``Autocatalytic reactions in the isothermal continuous
  stirred tank reactor,'' \emph{Chemical Engineering Science}, vol.~39, no.~6,
  pp. 1087--1097, 1984.

\bibitem{Pearson1993}
J.~E. Pearson, ``Complex patterns in a simple system,'' \emph{Science}, vol.
  261, no. 5118, pp. 189--192, 1993.

\bibitem{Iglesias2009}
P.~A. Iglesias and B.~P. Ingalls, \emph{Control theory and systems
  biology}.\hskip 1em plus 0.5em minus 0.4em\relax The MIT Press, 2009.

\bibitem{Kurtz1971}
T.~Kurtz, ``Limit theorems for sequences of jump markov processes approximating
  ordinary differential processes,'' \emph{Journal of Applied Probability},
  vol.~8, pp. 344--356, 1971.

\bibitem{Arnold1980}
L.~Arnold and M.~Theodosopulu, ``Deterministic limit of the stochastic model of
  chemical reactions with diffusion,'' \emph{Advances in Applied Probability},
  vol.~12, pp. 367--379, 1980.

\bibitem{Gillespie1977}
D.~T. Gillespie, ``Exact stochastic simulation of coupled chemical reactions,''
  \emph{Journal of Physical Chemistry}, vol.~81, no.~25, pp. 2340--2361, 1977.

\bibitem{Mazin1996}
W.~Mazin, K.~E. Rasmussen, E.~Mosekilde, P.~Borckmans, and G.~Dewel, ``Pattern
  formation in the bistable {Gray-Scott} model,'' \emph{Mathematics and
  Computers in Simulation}, vol.~40, pp. 371--396, 1996.

\end{thebibliography}

\end{document}